\documentclass[sort&compress]{aipproc}

\layoutstyle{8x11double}

\begin{document}

\title{Electroweak corrections to squark--anti-squark pair production at the LHC}


\author{W.~Hollik}{
  address={Max-Planck-Institut f\"ur Physik (Werner Heisenberg
  Institut) F\"ohringer
Ring 6, D-80805 M\"unchen, Germany}
}

\author{E.~Mirabella}{
  address={Max-Planck-Institut f\"ur Physik (Werner Heisenberg
  Institut) F\"ohringer
Ring 6, D-80805 M\"unchen, Germany}
}

\begin{abstract}
Presented are the complete NLO electroweak contributions 
to the production of diagonal squark--anti-squark pairs 
at the LHC.
We discuss their effects for the production of squarks 
of the first two generations, in different SUSY scenarios.
\end{abstract}

\maketitle


\section{Introduction}
If Supersymmetry (SUSY) is realized at the TeV-scale or below it will
be probed at the Large Hadron Collider (LHC). Experimental
studies  will be possible through the direct production of SUSY
particles. In particular colored particles will be copiously 
produced, so squark and gluino production can play an important role
in the hunting for SUSY. In the following we will focus on the production
of a squark--anti-squark pair,
\begin{equation}
P~P \to \tilde{Q}^a~\tilde{Q}^{a*}\, X\quad (\tilde{Q} \neq \tilde{t}, \tilde{b})\, . \,
\label{Eq:Process}
\end{equation}
The lowest order 
cross section for the process~(\ref{Eq:Process})
is of  $\mathcal{O}(\alpha_s^2)$ and was computed in the early 1980's~\cite{Tree,Tree2,Tree3,Tree4}. 
The dominant NLO corrections, of $\mathcal{O}(\alpha^3_s)$,
were calculated in Ref.~\cite{Beenakker1996}. They are positive and sizable, typically from 20\% to 30\%
of the lowest order prediction.\\
There are also $\mathcal{O}(\alpha_s \alpha)$ and $\mathcal{O}(\alpha^2)$
corrections to diagonal squark pair production from $q \bar{q}$ annihilation~\cite{Drees}.
Contributions of $\mathcal{O}(\alpha^2)$ are obtained squaring the tree-level EW graphs 
while $\mathcal{O}(\alpha_s \alpha)$ corrections arise from the interference of tree-level EW 
diagrams with the tree-level QCD ones. The latter vanish for $\tilde{Q}=\tilde{t}$, but they  
can become sizable if $\tilde{Q} \neq \tilde{t}$.\\
NLO electroweak (EW) contributions were found to be significant
in the case of top-squark pair production, with effects up to 20\%~\cite{StopEW, Beccaria:2007dt}. In the
case of the process~(\ref{Eq:Process})
NLO EW corrections can reach the same size as the tree-level EW contributions of $\mathcal{O}(\alpha_s\alpha)$
and $\mathcal{O}(\alpha^2)$~\cite{SquarkEW}.


\section{EW contributions}
Diagrams and  corresponding amplitudes for the EW contributions to the process~(\ref{Eq:Process})
are generated using \verb|FeynArts|~\cite{FeynArts,FeynArts2} while the algebraic manipulations and the numerical evaluation of the loop integrals 
are performed with the help of \verb|FormCalc| and \verb|LoopTools|~\cite{FormCalc,FormCalc2}.
IR and Collinear singularities are regularized within mass regularization.

\subsection{Tree level EW contributions}
Tree-level EW contributions to the process~(\ref{Eq:Process})  are of $\mathcal{O}(\alpha_s \alpha)$
and $\mathcal{O}(\alpha^2)$. The interference of the tree-level electroweak and tree-level QCD diagrams  give rise to terms of
order $\mathcal{O}(\alpha_s\alpha)$, while $\mathcal{O}(\alpha^2)$ terms are obtained squaring the aforementioned 
tree-level EW graphs. 
We consider  also the photon-induced partonic process  $\gamma g \to \tilde{Q}^a\tilde{Q}^{a *}$, which contributes 
at $\mathcal{O}(\alpha_s \alpha)$, owing to the
non-zero photon density in the proton which stems from the inclusion of NLO QED effects into the DGLAP equations 
for the parton distribution functions (PDFs).

\subsection{NLO EW contributions}
\begin{figure}
\includegraphics[height=.2\textheight]{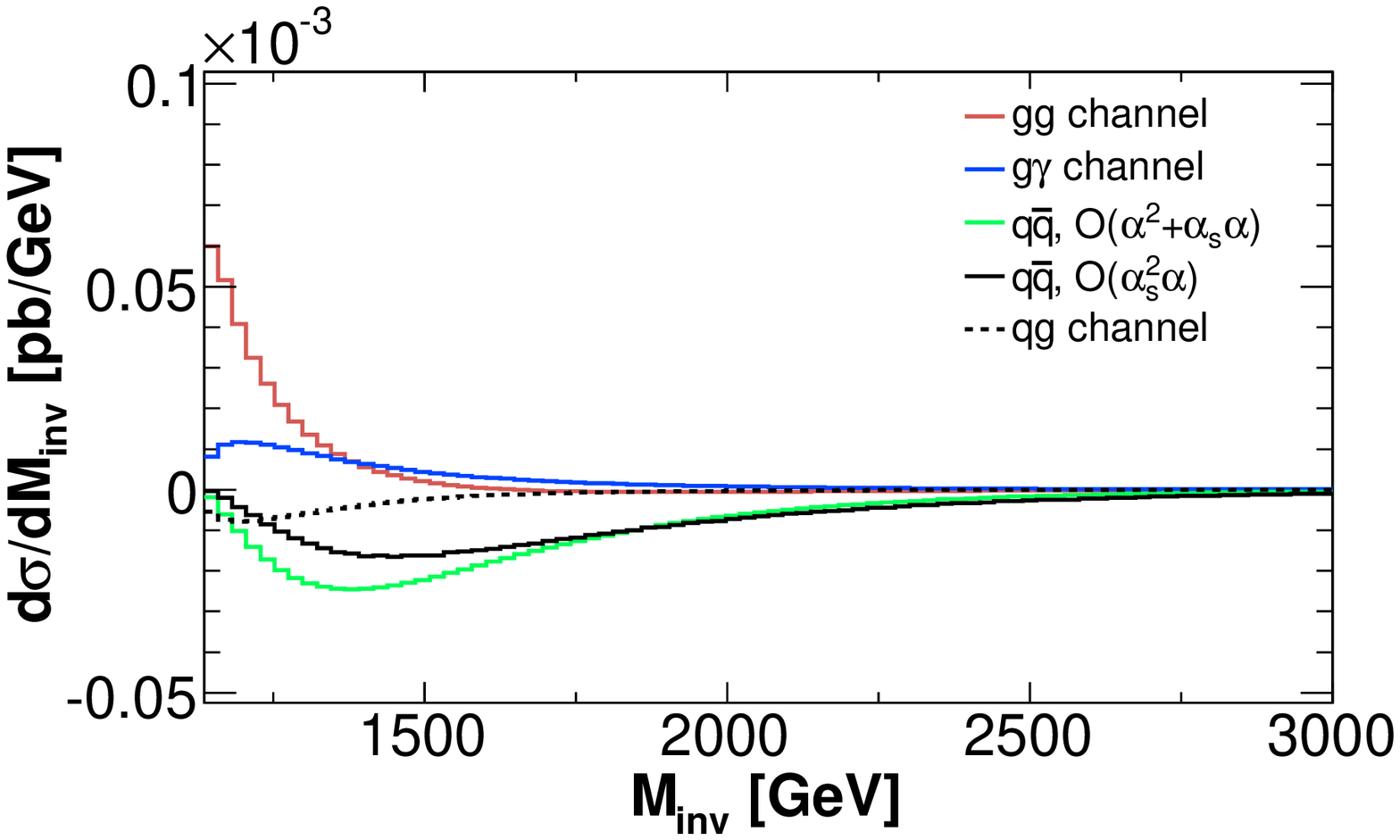}
\includegraphics[height=.2\textheight]{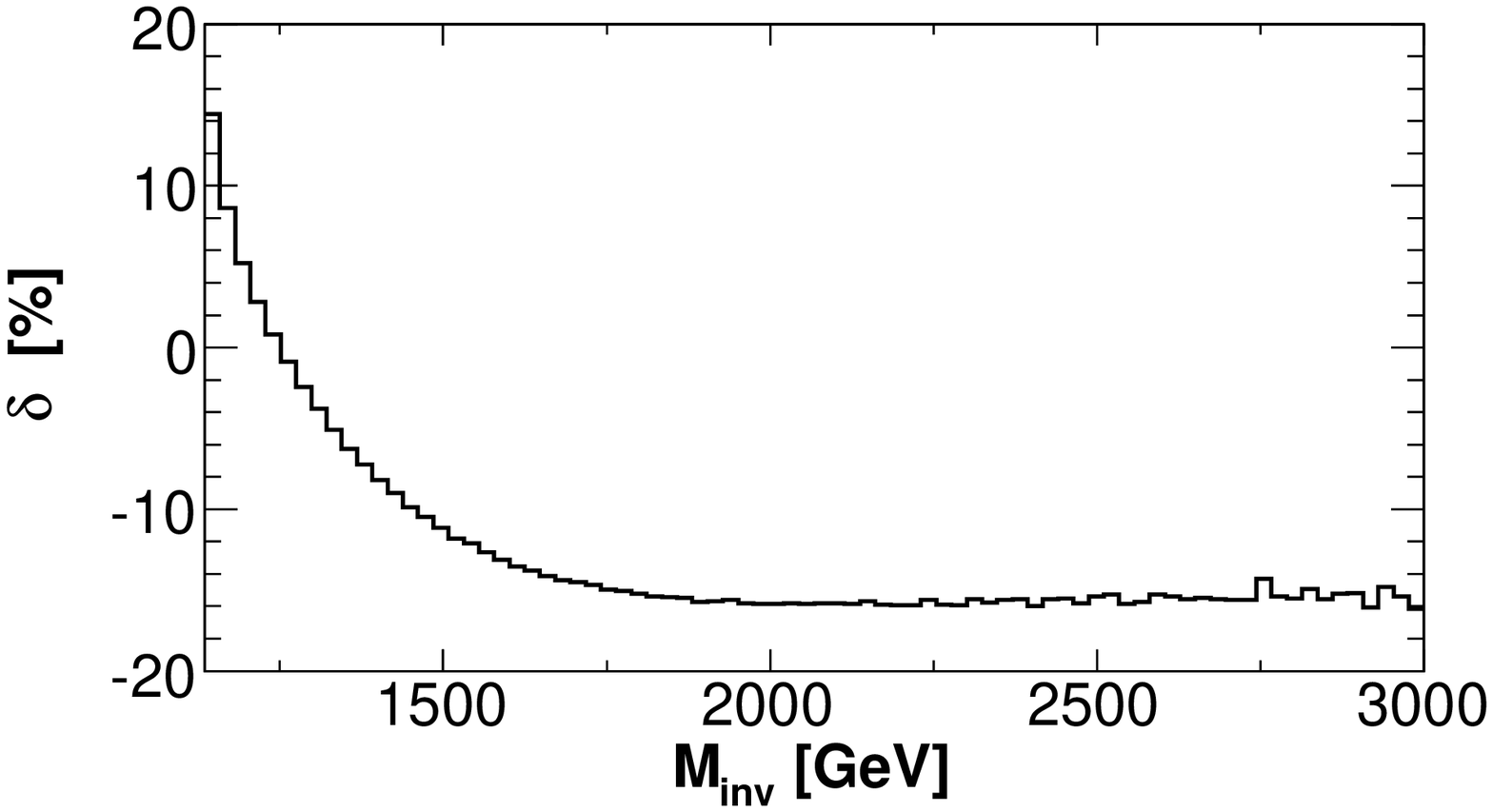} 
\label{Fig01}
\caption{Invariant mass distribution or $u^L u^{L*}$
production for the SUSY parameter  point corresponding to SPS1a$'$. The left panel shows the contributions
of the different channels. $\delta$ is the EW contribution relative to 
the LO one.}
\end{figure}
\begin{figure}
\includegraphics[height=.2\textheight]{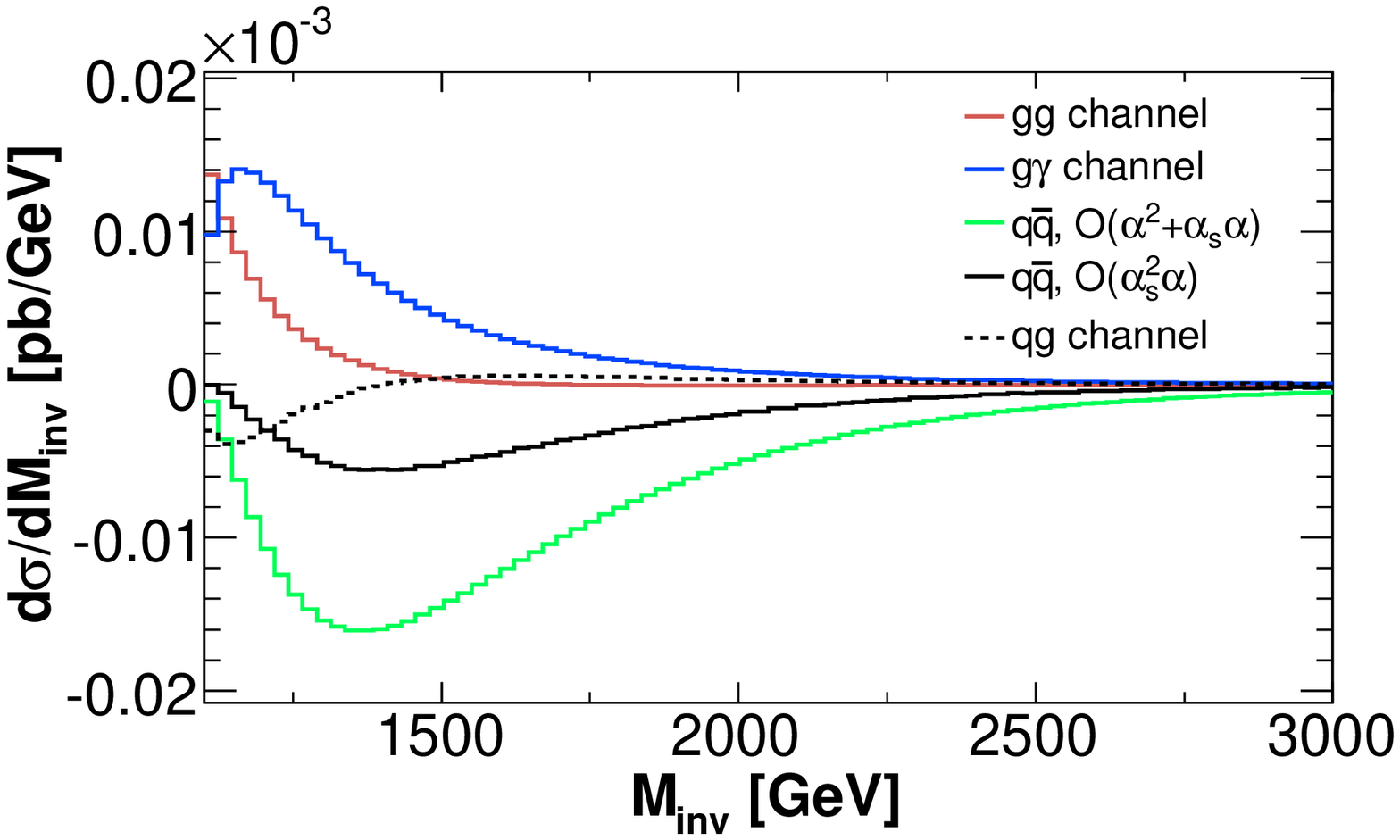}
\includegraphics[height=.2\textheight]{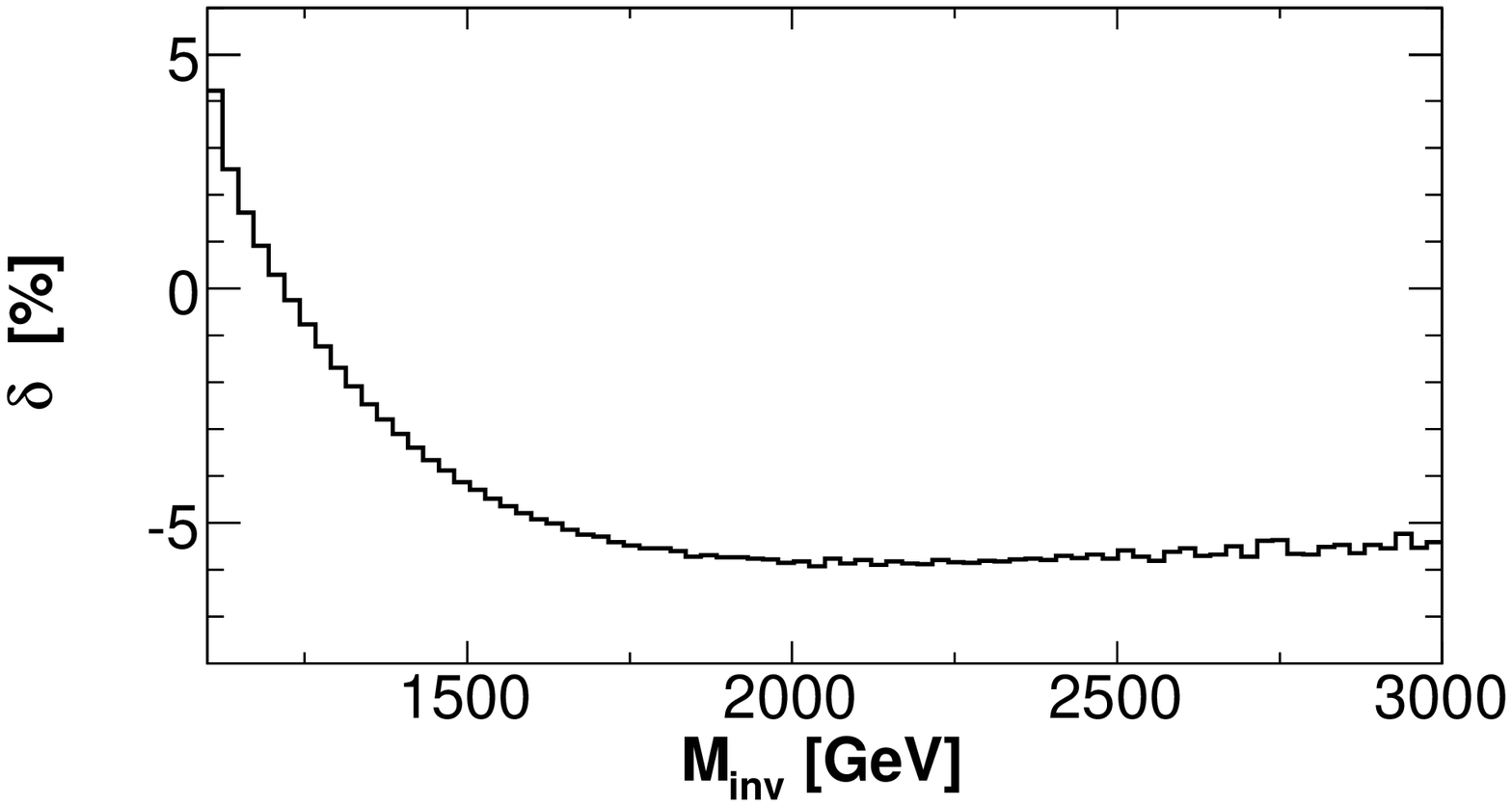}
\caption{Same as Fig.~\ref{Fig01} but for $u^R u^{R*}$ production}
\label{Fig02}
\end{figure}
NLO EW corrections arise from three different channels, gluon fusion, quark--anti-quark  annihilation,
and quark--gluon fusion channels.

\subsubsection{Virtual Corrections}
Virtual corrections originate  from the interference of the tree-level diagrams with the one-loop EW graphs.
In the case of $q \bar{q}$ annihilation channels, the interference between tree-level EW diagrams and QCD one loop graphs
has to be considered as well. Particularly interesting is the $q \bar{q}$ annihilation channel when the quark belongs to the same 
SU(2) doublet of the produced squark. In this case many types of interferences occur between amplitudes of 
$\mathcal{O}(\alpha_s\alpha)$ and $\mathcal{O}(\alpha_s)$ and between $\mathcal{O}(\alpha_s^2)$ and $\mathcal{O}(\alpha)$ amplitudes. This is 
related  
to the presence of EW tree-level diagrams with $t$-channel neutralino and chargino exchange and of QCD tree-level diagrams with $t$-channel
gluino exchange.  \\
The on-shell scheme~\cite{RzehakHollik,DennerHab} has been used to
renormalize masses and wavefunctions  of the squarks, of the quarks, and of the gluino.
The strong coupling $\alpha_s$ is renormalized in the
$\overline{\mbox{MS}}$ scheme. The contribution of the 
massive particles (top, squarks, and gluino) to the running of $\alpha_s$
has been subtracted at zero momentum transfer.
Dimensional regularization spoils SUSY at higher order; 
at one loop SUSY can be restored by adding a finite counterterm for the 
renormalization of the squark-quark-gluino  Yukawa coupling.

\subsubsection{Real Corrections}
IR and collinear finite results are obtained including
the processes of real photon emission. In the case of  $q \bar{q}$ annihilation
real gluon emission  of $\mathcal{O}(\alpha^2_s \alpha)$ has to be considered as well.
The treatement of IR and collinear divergences has been performed 
using two different methods: phase space slicing and dipole subtraction~\cite{Dipole}. They give 
results in good numerical agreement. \\
\noindent
IR singularities drop out  in the sum of virtual and real corrections. Surviving collinear singularities have to be 
factorized and absorbed into the definition
of the PDF of the quarks. \\

Other contributions  of  $\mathcal{O}(\alpha_s^2\alpha)$ arise from the 
processes of real quark emission from quark--gluon fusion. These contributions exhibit divergences
when the outgoing quark is emitted collinearly to the gluon. Such singularities are extracted using the two 
aforementioned methods and have been absorbed into the PDFs. In specific SUSY scenarios, the internal-state
gauginos can be on-shell. The poles in the resonant propagators are regularized 
introducing the width of the corresponding particle.


\section{Numerical Results}
\begin{figure}
\includegraphics[height=.2\textheight]{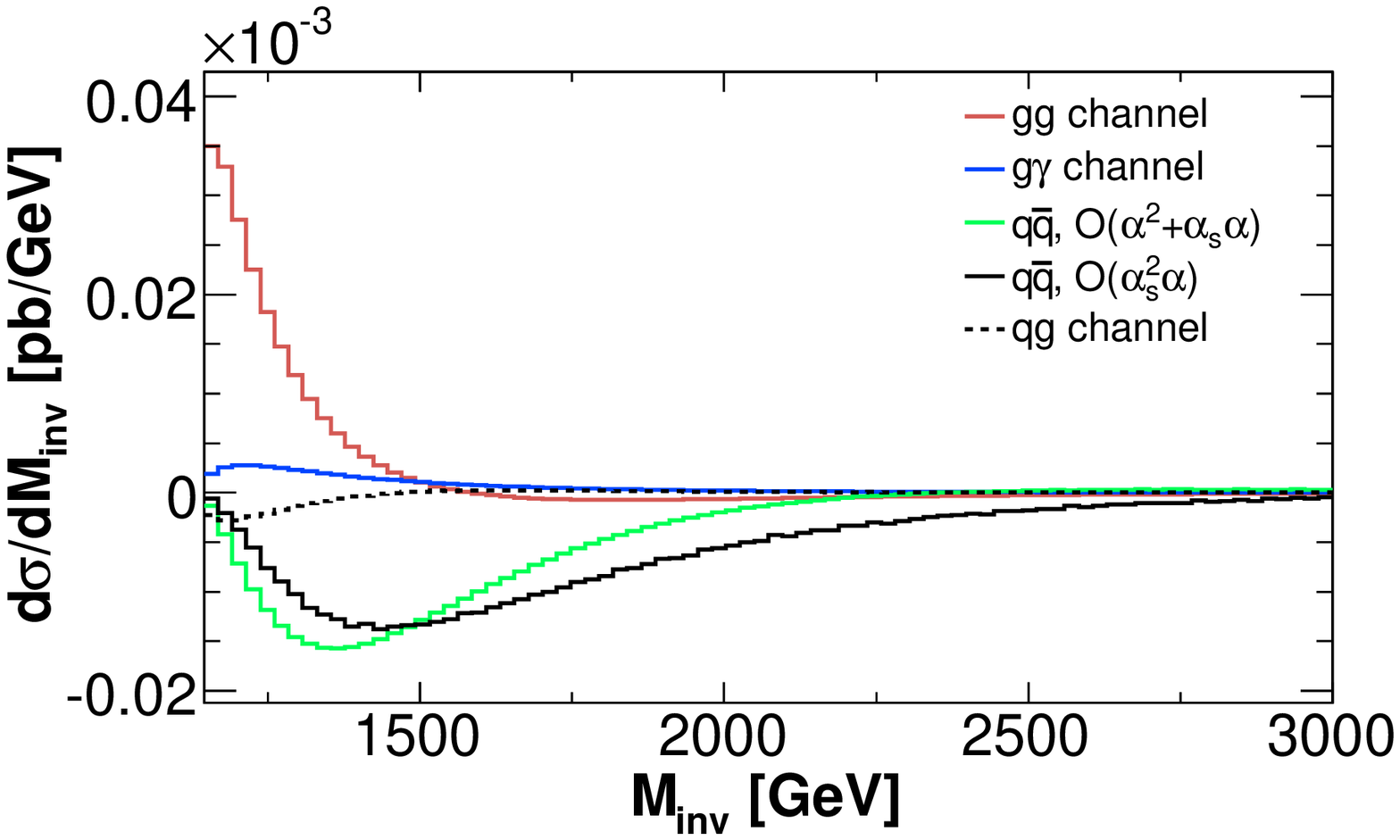}
\includegraphics[height=.2\textheight]{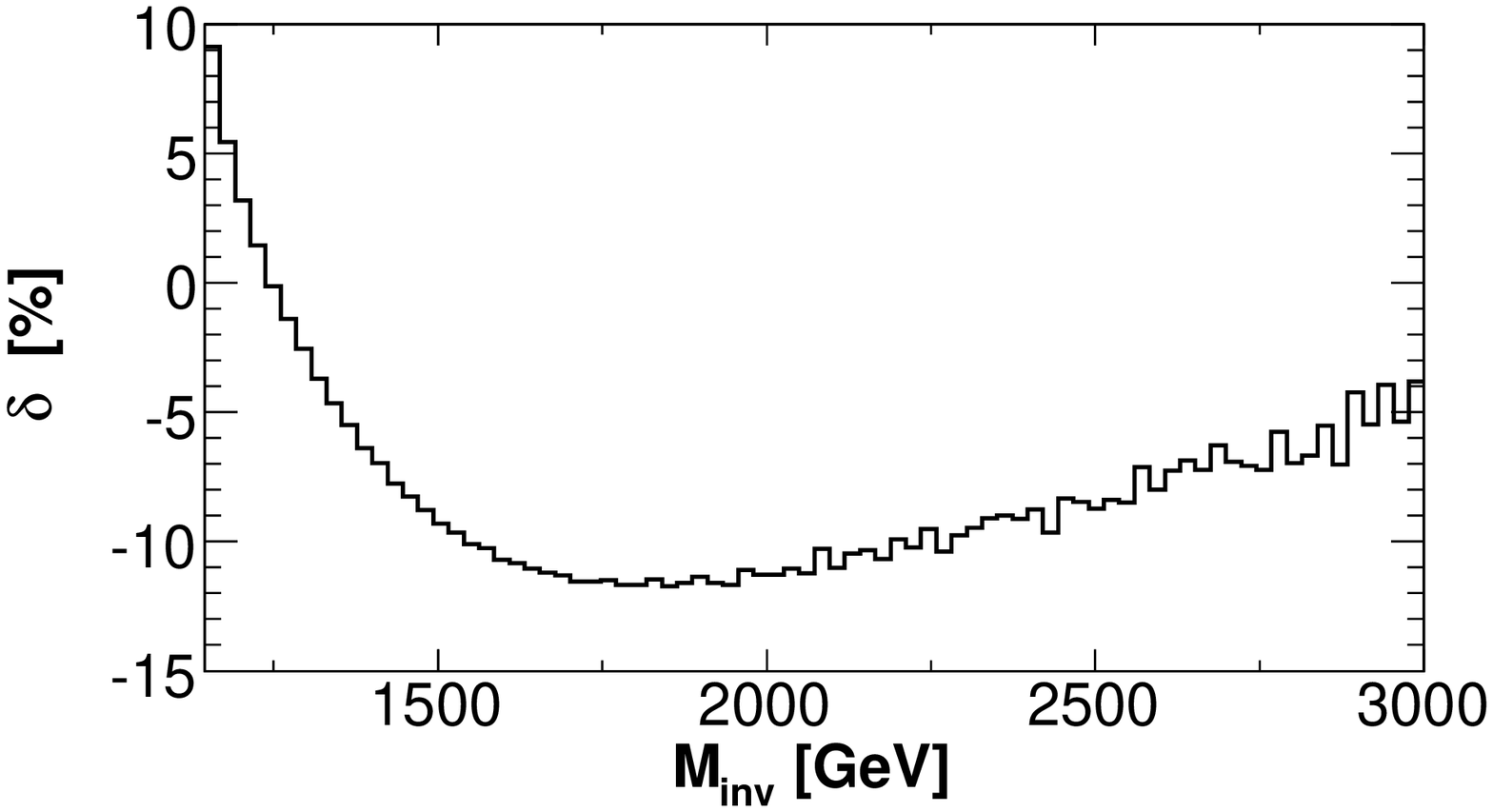}
\caption{Same as Fig.~\ref{Fig01} but for $d^L d^{L*}$ production}
\label{Fig03}
\end{figure}
\begin{figure}
\includegraphics[height=.2\textheight]{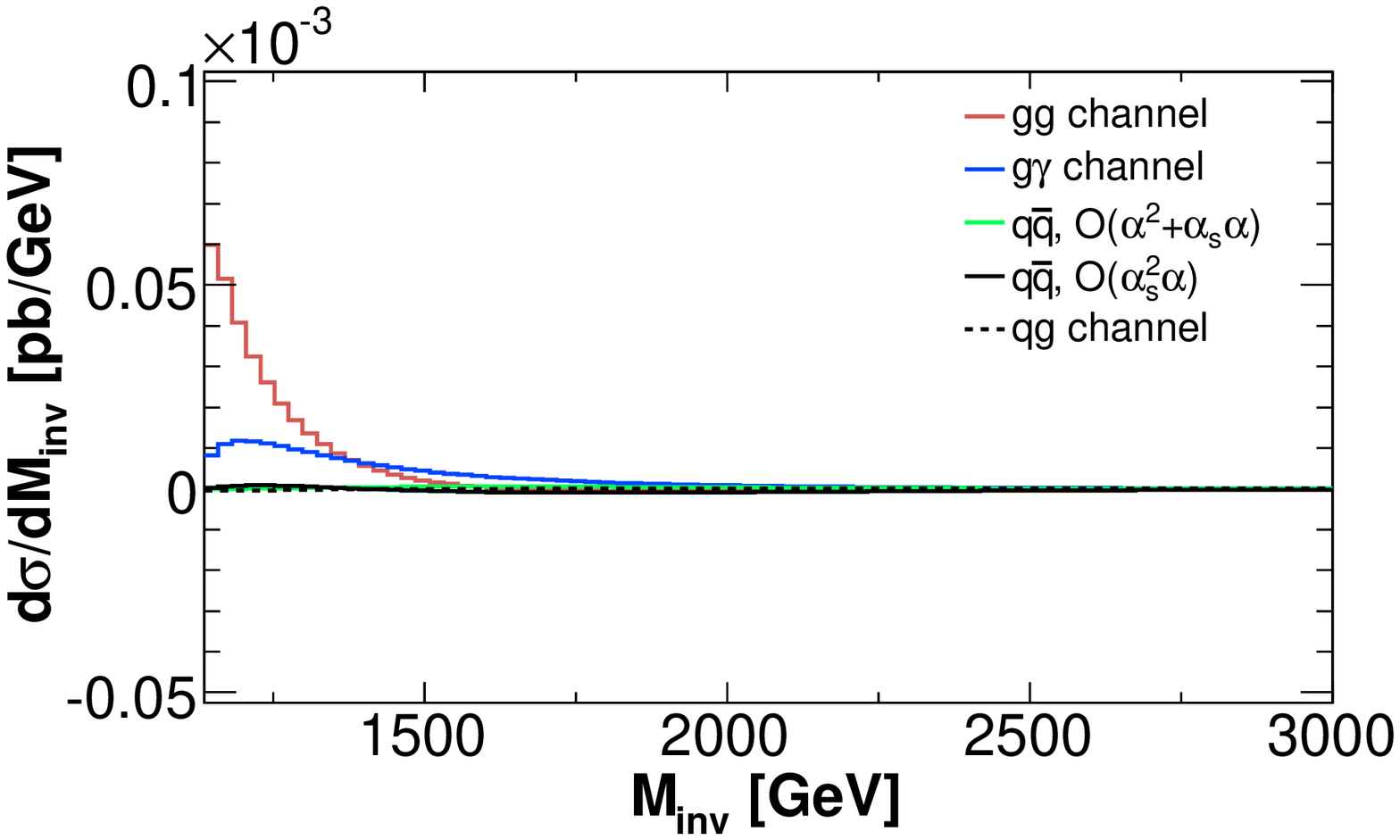}
\includegraphics[height=.2\textheight]{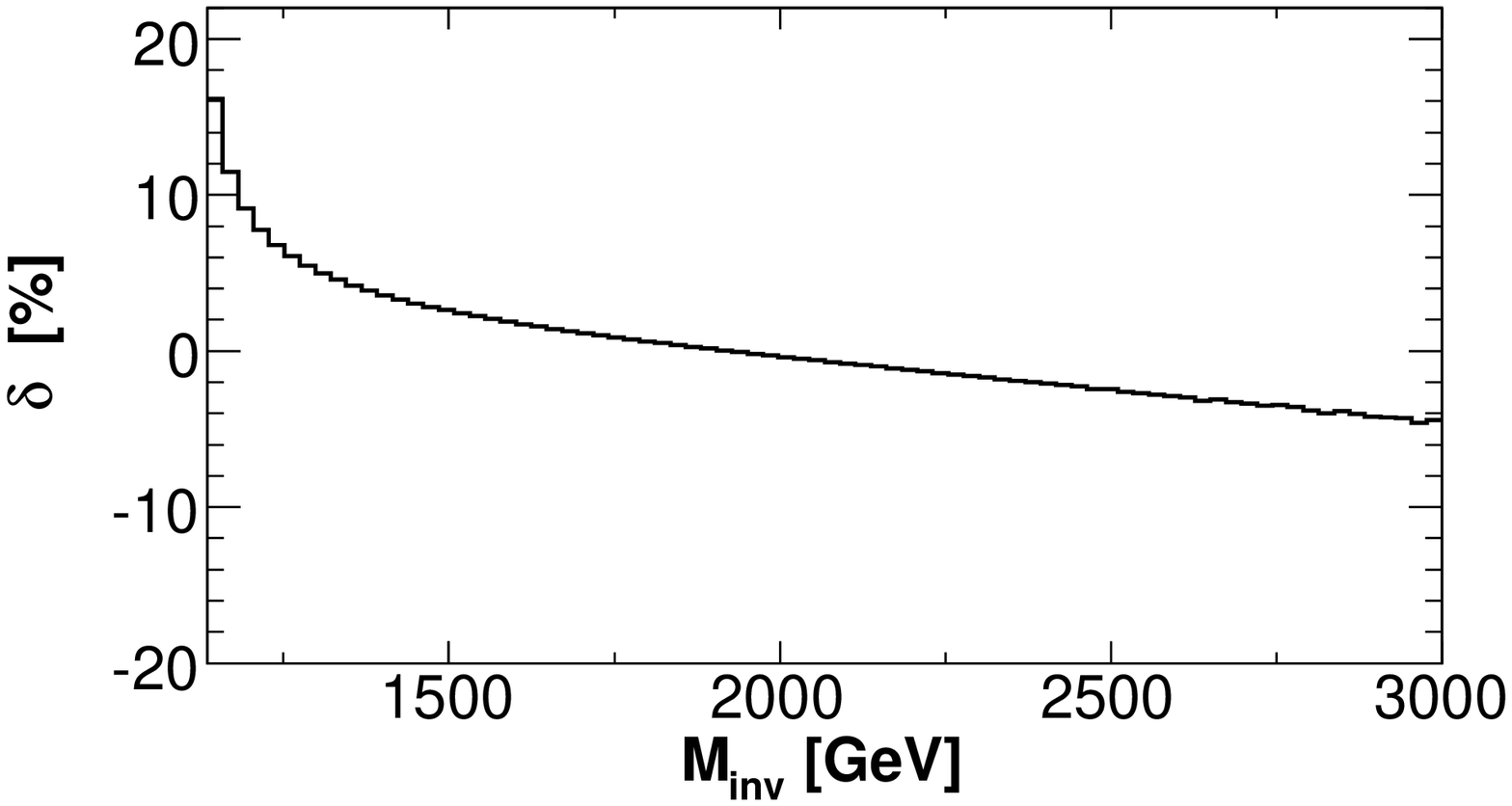}
\label{Fig04}
\caption{Same as Fig.~\ref{Fig01} but for $c^L c^{L*}$ production}
\end{figure}
For illustration of the EW effects, we study the pair production of the
squarks  $\tilde{u}^R$, $\tilde{u}^L$, $\tilde{d}^L$
and $\tilde{c}^L$, focusing on the SPS1a$'$ point of the MSSM parameter space,
suggested by the SPA convention~\cite{SPA}. A more comprehensive analysis
can be found in ref.~\cite{SquarkEW}. \\
\noindent
Figs.~\ref{Fig01}--\ref{Fig04} contain the invariant mass  distribution of the squark--anti-squark pair 
for the different squark species. In the low invariant mass region 
EW corrections are positive, decreasing as $M_{\mbox{\tiny inv}}$
increases and becoming negative in the high invariant mass region. The contribution of the $g\gamma$ channel 
is independent on the chirality of the produced squark, determined only by its charge. 
In the case of production of same-chirality and same-isospin squarks, {\it e.g.} $u^Lu^{L*}$ 
and $c^Lc^{L*}$,  the corresponding contributions of $gg$ and $g \gamma$ channels are equal ({\it c.f.} Fig.~\ref{Fig01} and 
Fig.~\ref{Fig04}), owing to the mass degeneracy of the produced squarks 
\footnote{Such degeneracy arises because quarks belonging to  the 
first two generations are treated as massless.}. 
Comparing Fig.~\ref{Fig01} and Fig.~\ref{Fig04}, one can understand the 
key role of the $q \bar{q}$  annihilation channels when 
the quark belongs to the same $SU(2)$ doublet of the produced squark. In the case of $u^Lu^{L*}$ production the contribution of these channels is  
negative and sizeable 
while in the case of $c^L c^{L*}$ production it is suppressed by the PDFs rendering the impact of the $q \bar q$ channels negligible.


\bibliographystyle{aipproc}   
\bibliography{ref}

\end{document}